# Macros and Multiscale Dynamics in Spin Glasses

N. Persky and S. Solomon *
Racah Institute of Physics, Hebrew University, Jerusalem 91904, Israel

We study the spacial and temporal multiscale properties of complex systems. We present accelerated algorithms for dilute spin glasses and display explicitly their relation to the effective dynamics of specific collective degrees of freedom (macros).

We discuss the difficulties in applying multiscale-cluster algorithms (MCA) to general frustrated systems: MCA does not succeed to break the system into clusters. We relate these difficulties to rigorous negative results in systems with an ultrametric space of ground states: the tunneling between vacua cannot be expressed into an algorithm acting upon independent macros.

## I. COMPLEX SYSTEMS AND MACROS

One of the main characteristics of complex systems is their computational difficulty: the time necessary for their investigation and/or simulation grows very fast with their size [1]. The systematic classification of the the difficulty and complexity of computational tasks is a classical problem in computer science [2].

In physical systems, the emergence of large time scales is often related to multiscale spacial structures within the system. Long range and long times scale hierarchies (**Multiscale Slowing Down**) are usually related to collective degrees of freedom (**macros**) characterizing the effective dynamics at each scale.

The physical understanding of complex macroscopic phenomena is then often expressed through identifying the relevant **macros** and their effective dynamics (e.g. hadrons in the theory of quarks and gluons, Cooper pairs in superconductors, phonons in crystals, vortices in superfluids, flux tubes, instantons, solitons and monopoles in gauge theories, etc.).

One can entertain the hope that many complex systems display some kind of universal multiscaling exponents generalizing the scaling critical exponents of the critical systems. One could hope for the existence of some kind of Multiscale-universality classes generalizing the universality classes of Renormalization Group theory. Such a situation would have a significant unifying effect on a very wide range of phenomena spreading over most of the contemporary scientific fields.

In the absence of a rigorous theoretical basis for such a hope, its investigation relies for the moment mainly on the use of computers. In particular one uses "first principles" simulations which implement directly and without the intermediary of ad-hoc approximations the fundamental physics of the systems under study.

Usually, it is the dynamics of the **macros** during simulations which produces the Multiscale Slowing Down and reciprocally, the slow modes of the simulation dynamics project out the relevant macros [12].

Therefore, a better theoretical understanding of the multiscale structure of the system, enables one to construct better algorithms by acting directly on the relevant macros. Reciprocally, understanding the success of a certain algorithm yields a deeper knowledge of the relevant degrees of freedom of the system [1].

The present paper implements this point of view into the study of spin glasses.
Section 2 introduces the basic notions of Multiscale-Cluster Algorithms (**MCA**).
Section 3 describes the difficulties in applying MCA to generic frustrated systems.
Section 4 contains rigorous results which forbid macros in ultrametric (UM) systems.
Section 5 identifies the relevant macros and their role in constructing MCA for Dilute Spin Glasses. Section 6 demonstrates numerically the efficiency of the resulting MCA.
Section 7 summarizes the conclusions. Appendix A contains the proofs of the results stated in Section 4.

We interpret the negative results in the pair of sections 3-4 and the positive results in the sections 5-6 as supporting in both directions the relation between macros and the efficiency of MCA.

## II. MULTISCALE-CLUSTER ALGORITHMS (MCA)

An example of multiscale effective dynamics and its related multiscale slowing down is the critical slowing down (**CSD**) at second order phase transitions. There, the relaxation time $\tau$ diverges with the systems size $L$ as:

$$\tau \sim L^z$$

*Email: nathanp@vms.huji.ac.il sorin@vms.huji.ac.il

[1]see for example of the projection by PTMG of exact lattice Atyiah-Singer modes [8].

where $z$ ($\sim 2$) is the dynamical critical exponent. Consequently, the typical time needed to produce a large Boltzmann set of decorrelated configurations diverges and the standard local Monte-Carlo methods become inefficient.

It was shown that when the detailed knowledge on the relevant macros is included in the simulation algorithms, the value of $z$ can be reduced dramatically (down to 0) [28]. These algorithms, which we will call generically here Multiscale-Cluster Algorithms (MCA), allow very fast and precise computation of the equilibrium thermodynamic properties of the systems. However their main importance is to guide and validate by objective means (lowering of z) the intuitive guesses on the physically relevant macros and their macroscopic dynamics [12].

We are treating the various (**MCA**) in a conceptually unified way: as expressions of the macros appearing at various scales. In fact many of the explanations in Section 5 on the dynamical relevance of the spatial structures manipulated directly by the Macros Reduction Algorithm (MRA) can be given equally in the language of Dynamical Algebraic Multigrid (DAMG) [9] as well as in the framework of the Cluster Algorithm (CA) [5].

In statistical mechanics systems, the objective of MCA is to generate as fast as possible a representative sample of configurations. This is realized by acting directly on the macroscopically relevant macros (in contrast with the usual local algorithms which act on the microscopic elementary degrees of freedom).

The typical cluster algorithm (CA) works according the following general principles:

1. One selects a particular subset of allowed changes for the degrees of freedom associated with each site $i = 1, .., N$ of the system [2].

   This reduces the system to an Ising-like system $S = \{s_1, ..., s_N\}$ where the Ising variables $s_k$ can take the values $\pm 1$. A configuration is a specific assignment of one of these values for each $s_k$. The interaction energy

   $$E(S) = \frac{1}{2}\sum_{i,j} J_{i,j}(1 - s_i s_j).$$

   is parametrized by the **link** parameters $J_{i,j}$ associated with each pair of sites $i$ and $j$ [3].

2. One constructs and updates a system of clusters which preserves the macroscopic dynamical properties of the initial spin system.

   The system of clusters and its dynamics is obtained by modifying the (**link**) parameters $J_{i,j}$ between the pairs of spins $(i, j)$. More precisely the link $(i, j)$ is either "frozen" $J_{i,j} = \infty$ or "deleted" $J_{i,j} = 0$ based on the following classifications.

   Consider the current values $s_i^C$ and $s_j^C$ of the 2 spins and their current energy: $E(S_{i,j}^C) = \frac{1}{2}J_{i,j}(1 - s_i^C s_j^C)$.

   - If the spins are in the low energy state $E(S_{i,j}^C) = 0$. the link $(i, j)$ is called saturated or **satisfied**. Otherwise, the link is said **unsatisfied**.
   - If the difference between the satisfied and unsatisfied energies of the link $|J_{i,j}|$ is big, the link is called **strong**. Otherwise it is **weak**. Often the strong-weak label is given relative to the actual temperature of the system. For instance at very large temperature ($T >> |J_{i,j}^C|$) all the links can be considered weak while at low temperature ($T << |J_{i,j}^C)|$) most of the links may be acting as strong.

   With this terminology, the cluster generating procedure is:

   (a) freeze (with high probability) the strong satisfied links (links with low energy).

   (b) delete the strong unsatisfied links (links with high energy).

   (c) give for weak links an appropriate stochastic chance to both options (frozen-deletion) to arise.

   (d) flip the relative signs of spins which belong (by the link deletions) to different clusters.

In the sequel we will call **loop** a closed chain of links

$$\{(i_k, i_{k+1}) \mid k = 1, ..., n \; and \; i_{n+1} \equiv i_1\}.$$

If the product

$$\prod_{k=1,n} J_{i_k, i_{k+1}} < 0$$

is negative the loop is said **frustrated**. If a loop is frustrated there exists no spin configuration for which all the links of the loop are satisfied.

[2] E.g. in finite temperature SU(2) gauge theory, the SU(2) matrix degree of freedom on the time-like links are allowed to change only their signs during a MCA step [10]. This is an algorithmic expression of the physical understanding that it is the center of the group which is the relevant degree of freedom.

[3] For notational convenience, we use in this and the following section a definition of the total energy which differs by an overall additional constant $E_0 = \frac{1}{2}\sum_{i,j} J_{i,j}$ from the definition used in the rest of the paper.

## III. MCA DIFFICULTIES IN FRUSTRATED SYSTEMS

The problem of the applicability of MCA to frustrated systems arised quite early since most of the cases in which the MCA **did not work** were situations in which the first step of section 2 reduced the system to a frustrated one [3,12]

Some of the most important families of frustrated systems are the Randomly Frustrated Systems (RFS) and the Spin Glasses (SG).

A typical SG system presents a complex energy landscape consisting of many local minima, separated by huge barriers which scale with the size of the system. This is expressed by the emergence of an ultrametric structure of the ground states space and an infinite hierarchy of exponentially divergent relaxation times [13].

To understand the difficulties which occur when applying MCA to frustrated systems let us analyse in detail a simple scenario. Suppose that the configuration (C1) in the following figure:

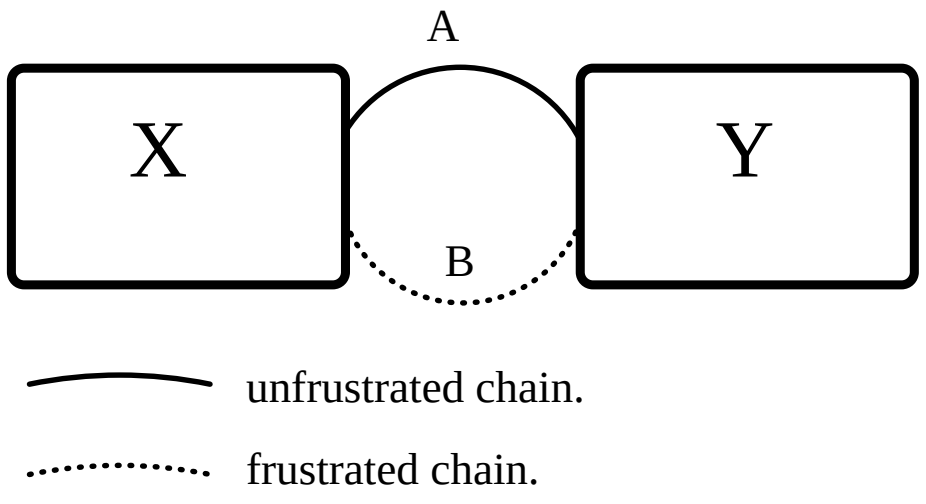

is an energy ground state (GS). Consider the case that the subsystems X and Y are linked by 2 chains of links A and B (as shown in the figure).

Suppose that in C1, the total energy of the links belonging to A is $E_A^1 = 0$ while the energy of the chain B is is $E_B^1 = e$. The chain B is therefore frustrated while the chain A is unfrustrated. For the simplicity of the argument let as assume that the system is at very low temperature, $T << e$ though the conclusion is valid at higher temperatures ($T < e$) as well.

Assume now that a new ground state (GS) configuration C2 is obtained by flipping in C1 the relative sign of X and Y such that $E_B^2 = 0$ and $E_A^2 = E_B^1 = e$. This means that in the ground state C2 it is the chain A which is frustrated while B is unfrustrated. The total energy of C2 is equal to the total energy of C1. An efficient algorithm should allow one to easily obtain C2 from C1 and viceversa.

However, with the usual cluster algorithm, this condition is not fulfilled. Indeed, if the chain A is frustrated (in C2) so that the chain eventually would be cut by the Cluster Algorithm (CA) then, automatically B is unfrustrated and all its links will be eventually frozen by the CA. Vice-versa: if B is cut, A is uncut (this happens in C1). Consequently, in any case, at least one of the chains A and B is frozen. Therefore X and Y are always included by the CA in the same cluster and there is no way to get from C1 to C2.

Therefore it appears that a frustrated disordered system is not compatible with an efficient updating algorithm. This argument is getting even stronger if X and Y are linked by several chains. We will present in section 4 some rigorous results extending these intuitions to a wider class of systems.

In a few special cases one can overcome these problems. If the problem is local and the structure of the links is completely known, then a "two bond deletion" might help such as in the fully frustrated system on a square lattice [14]. The "two bond deletion" technique can be extended in other systems to 3 and 4 bond deletions [24] and even to an "n"-bond procedure. However, in general, the problem would still revert to exponentially combinatorial complexity if one has no a-priory knowledge about which "special" subset of links one should combine.

The simulated annealing technique which helped some systems from getting stuck in local minima, failed to provide a complete solution in SG case. A related direction is the SW replica algorithm [5] and its modifications [23] which use simultaneously various replica of the system in order to identify large spatial regions which act coherently. This might work in few simple cases, in low dimension (until now only 2D). However, one cannot expect such an algorithm to work for a general frustrated system because:

- in general (e.g. for spin glasses [4]) one can find for each GS exponentially many metastable states with energy close to the ground-states. As a result one needs in general an exponential number of replica in order to fully capture the structure of the system.
- The spacial regions which have to flip in order to turn various ground-states one into another cannot in general be identified and manipulated as independent entities (see in Section 4 the relevant theorems for ultrametric systems).

These limitations imply that one cannot get away from the combinatorial complexity of the general randomly frustrated problem. The CA logic of constructing clusters, is based on **local** (see e.g. [27]) features implying **local** criteria such as deleting the unsatisfied links which define the block boundaries in the ferromagnet case. However in a general case the feature of the cluster is only apparent in a **global** view, without any local signs (see theorem III).

The situation can be compared with having to find one's way in a labyrinth in the phase space: each small local change in the position of the potential energy labyrinth walls determines large unpredictable changes of the solution route depending on details scattered across the entire phase space.

Consequently, we are discerning 3 main complexity cases:

- In very simple cases the pattern of complexity is reducible (may be by an iterative multiscale procedure) and the MCA's capturing this reducible complexity are an efficient computational and conceptual tool.
- In the general case one has to put an exponential computational effort to fully "understand" the structure of the system. This situation is similar with "understanding" the architecture of a labyrinth and is expressed by the theorems of Section 4.
- In some cases the system contains certain macros which are "irreducibly complex". Yet the interactions between these macros are tractable by MCA or other algorithms. In these cases, MCA can help reduce the "less complex" part of the dynamics leaving the "irreducible core" for a separate treatment.

The last possibility has been exploited in the Parallel Transported Multigrid (PTMG) [11,8] treatment of the fermions in gauge field background where the complexity related to the gauge freedom was eliminated at the MG level while for macros related to frustration and topology (e.g. Atyiah-Singer zero-modes) one has developed a method [12] for implicit identification, manipulation and elimination of irreducibly complex macros. Similar intuitions are at the basis of the successful algorithms for diluted spin glasses described in sections 5 and 6.

Recognizing the "irreducibly complex" parts of a complex system (rather than trying vainly to solve them by multiscale means) might be a very important aspect both conceptually and computationally.

## IV. ULTRAMETRIC (UM) SYSTEMS DO NOT HAVE INDEPENDENT MACROS

As explained above, the SG systems present a certain hierarchy in their energy landscape which is responsible for the hierarchy of time scales characterizing their multiscale slowing down. This rugged energy landscape is also the origin of the ultrametric properties of their ground states space.

One could hope to make some relation between the SG ultrametric hierarchy and the existence of an effective representations of the dynamics in terms of a spacial multiscale hierarchy of independent macros. This in turn would become the basis of an efficient MCA.

It turns out that the case is exactly the opposite: the ultrametric hierarchy characterizing SG **insures** the **in**existence of a representation of the effective macroscopic dynamics of the complex system in terms of their macroscopic disjoint sub-sets (i.e. a complex ultrametric system is not effectively reducible to a set of subsystems).

Let us prove it in detail.

Consider an Ising-like system with variables $s_k = \pm 1,\ k = 1, ..., N$. The energy of a configuration $S^C = \{s_1^C, ..., s_N^C\}$ is $E(S^C) = -\frac{1}{2}\sum_{i,j} J_{i,j} s_i^C s_j^C$. The ground states (GS) of $S$ are the configurations with the lowest energy density (states differing only by a finite energy ($O(N^0)$) are considered degenerate).

The metric in the configurations space is defined by the following distance. If 2 configurations $S^0$ and $S^A$ differ (only) by the sign of the spins belonging to a subset $A$ then their distance is

$$d(S^0, S^A) = \#A = \textit{the number of elements in } A. \quad (1)$$

Configurations differing only by a global change of sign are considered identical. This brings the support of the function $d(S^A, S^B)$ into the interval $[0, 0.5N]$.

We shall call a system ultrametric (**UM**) if the GS's with the above metric form an ultrametric space. Namely, for any 3 GS's $S^C$ , $S^A$, $S^B$ one has

$$d(S^A, S^B) \leq max[d(S^C, S^A), d(S^C, S^B)]. \quad (2)$$

Note that for real systems this condition is fulfilled, probably up to measure zero of violations, and up to some small $\epsilon \sim d/N$ [18]. Those limitations are not affecting our final conclusions though one should be aware to their existence.

The first theorem expresses the fact that one cannot hope to travel between various GS's of a ultrametric system by just identifying and flipping independently various subsets (macros/collective objects).

More precisely:

**Theorem I**:

• Consider 2 subsets $A$ and $B$ of of a ultrametric system $S$ which has a GS $S^0$. Assume that the states $S^A$ and $S^B$ obtained by flipping (only) the spins of the set $A$ (respectively $B$) in $S^0$ are GS's too.

**then**

• For $\#A = \#B$

$$\#(A \bigcap B) \geq 1/2 \#A \quad (3)$$

• For $\#A \neq \#B$

$$\#(A \bigcap B) = 1/2 min(\#A, \#B) \quad (4)$$

Theorem I means that at least half of the spins of one of the sets ($A$ or $B$) belong to the other set. This is hardly ones idea of two independent sets. Moreover, generally, a point belongs to an infinite number of strongly overlapping clusters. This implies that **locally** one has no criterion for constructing the relevant macros. Those can be identified only from a **global** view.

In conclusion, in ultrametric systems it is ruled out that various regions of the system can be treated as independent collective degrees of freedom (**macros**). This picture can be extended to finite but small temperatures with the help of the "pure state" concept [1].

This failure of separability of the whole into (almost) independent parts has conceptual implications in the sense that one cannot "understand" the complex system by "analyzing" it into its parts. In this sense an ultrametric system is conceptually irreducible to simpler entities. We will see in the next section that optimal global algorithms reduce in fact a system to its "irreducible" core.

One is tempted to conclude that the entire discussion of reductionism can be reformulated in terms of "irreducible complex systems". I.e. in place of **assuming** ultrametricity and deducing the inexistence of independent dynamical sub-objects, one can propose this **dynamical inseparability** as the fundamental property underlying irreducible complexity.

*Theorem I* suggests therefore that one should engage in the systematic study of the systems which have families of GS's differing by strongly overlapping subsets. The topology induced in the system by these subsets might have interesting properties.

In the sequel of this section, we will give further characterization of the sets $A$ which relate (by their flipping) different ground states.

**Theorem II**:

• If $S^0$ and $S^A$ differing (only) by the sign of the spins in the region $A$ are both GS of a (not necessarily ultrametric) system S,

**then**

• their actual spin arrangement restricted to the system A alone is a GS of the system A considered as isolated from the rest of S.

Note that this statement holds **not only** in low dimensions (where the surface energy is not extensive). To see in which respect this statement is non-trivial, note that in the presence of the system $\bar{A}$ (the complement of $A$ in $S$), the spins in the system A are submited to the influence of the external (to $A$) action of the spins in $\bar{A}$. For a general subset $A$ of $S$, this will bring the spins of $A$ into positions which are not necessarily optimal in terms of the internal $A$ interactions alone. They would be in general in a position which strikes a compromise between minimizing the internal $A$ energy and the interactions with the rest of the system ($\bar{A}$). Theorem II finds conditions in which the action of $\bar{A}$ can be ignored.

This property has interesting uniqueness consequences on the sets flipping between GS's of ultrametric systems.

**Theorem III**:

• If $S^0$ and $S^A$ differing by (and only by) the signs of the spins in $A$ are GS's in the ultrametric system $S$.

**then**

• their actual spin arrangement restricted to the system A alone, is **the unique** GS of A considered as isolated from the rest of S.

This theorem throws some ironical light on the properties of GS's in ultrametric systems: *a posteriori* **there is** something qualitatively special in the sets which connect between GS's of UM systems: the uniqueness of the **their** GS. These sets are very special and do not share at all the proliferation of vacua characteristic to typical SG systems. In fact theses subsets are not UM systems by themselves: Theorem III implies (among other things that) these subsets cannot constitute a hierarchy of **UM** sub-systems included recurrently one into the other.

## V. MACROS REDUCTION ALGORITHM (MRA) IN DILUTE SPIN GLASSES

The "no-go" arguments of section 3 had a rather depressing effect on the expectations of the practitioners in the field on the performance of MCA in frustrated systems. We will see below that these arguments and even the theorems presented in section 4 still allow for a significant contribution of MCA in frustrated systems as long as they possess macros.

As opposed to fully connected models such as SK [21], the geometry of the diluted models includes topological structures capable to engender such macros. The cluster algorithms (CA) can then locate and act on large regions of the configuration which are weakly linked to the rest of it. In addition to CA we construct a Macros Reduction Algorithm (**MRA**) which acts **explicitly** on the same macros on which CA acts stochastically. In this way we make explicit the role of the macros in both algorithms. The MRA has a structure very similar to the Dynamical Algebraic MultiGrid (DAMG) of [9]. Since its action is more direct, MRA is more efficient than CA in the models for which it was designed (see Section 6 for numerical details). However CA is more versatile.

We consider again the Ising-like system

$$H = -\frac{1}{2}\sum_{i\neq j} J_{ij} s_i s_j \tag{5}$$

with the probability distribution for $J_{i,j}$:

$$P(J_{ij}) = (1 - c/N)\delta(J_{ij}) + (c/N)f(J_{ij}). \tag{6}$$

$f(J_{ij})$ is the distribution of the surviving links after the dilution.

This model, is known as the highly diluted system SG with finite average connectivity $c(= O(1))$. The probability for a spin to have connectivity $k$ in such a system follows the Poisson distribution: $c^k \exp(-c)/k!$.

Many geometric properties of this system are well understood [19,20]. In particular, the system undergoes a percolation transition at $c = 1$. The maximal cluster is of order $O(logN)$ for $c < 1$, $O(N^{2/3})$ at $c = 1$, and $O(N)$ for $c > 1$ where its size is explicitly given by $P = 1 - P_0 = 1 - \exp(-cP)$.

The finite connectivity models at low temperature are directly connected to the graph partitioning problem [19,20] (dividing a graph into subgraphs, with minimum connections between them).

### A. The Macro Reduction Algorithm

In the introduction, it was claimed that the very existence of an efficient MCA may help identify relevant macros in the system.

To achieve this, we construct a Macros Reduction Algorithm (MRA) which freezes explicitly spins into macros and by doing so reproduces (and improves over) the performance of the efficient Cluster Algorithm (CA).

MRA consists of the following iterative steps

1. Access the points $i$ of the system iteratively starting preferably with the ones with lower connectivity.

2. For an accessed point $i$ find its strongest connection $J_{i,j}$ defined by $|J_{i,j}| > |J_{i,k}|$ , *for all* $k \neq j$.

3. Freeze $s_i$ and $s_j$ into a macro such that $-s_i J_{i,j} s_j$ is minimal. From now on the value of $s_j$ alone labels the state of the macro and $s_i$ is just a "slave" determined by it.

4. Ignore $s_i$ in the subsequent updating's of the system (in particular in the updating's of $s_j$).

5. As seen in the detailed explanations below, when all points (macros) with 1 neighbour are exhausted, all trees are shrink to points.

6. When all points (macros) with 2 neighbours are exhausted, all linear chains in the system shrink to length 1.

7. After that, some regular MC (either local or global) which acts on the new system can be used.

Note the the "reduction" stage takes negligible time (few MC steps) compared to the relaxation part!.

The construction of macros in MRA is very similar to the block construction in Dynamical Algebraic MultiGrid (DAMG) described in [9].

Let us see now in detail why MCA's such as CA, AMG and MRA, work where the local algorithm doesn't.

#### B. trees

Consider a configuration:

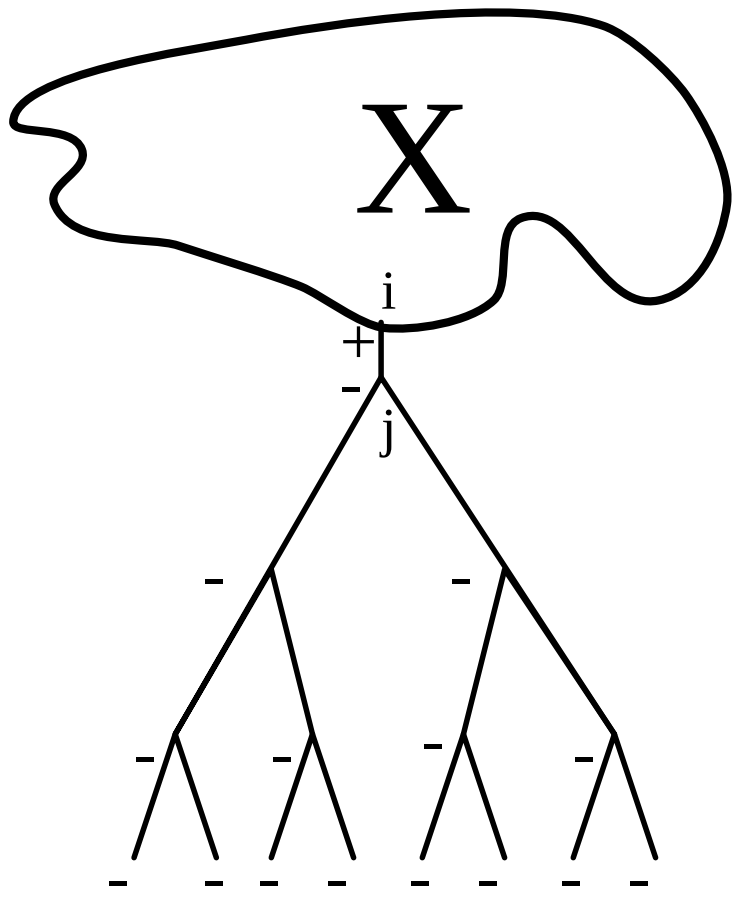

Assume X is minimal if $s_i = +$. Then the energy is minimized if all $s$'s on the tree sites are $+$. However , a simple MC algorithm might have problems in reaching this minimal energy configuration. For instance if one has:

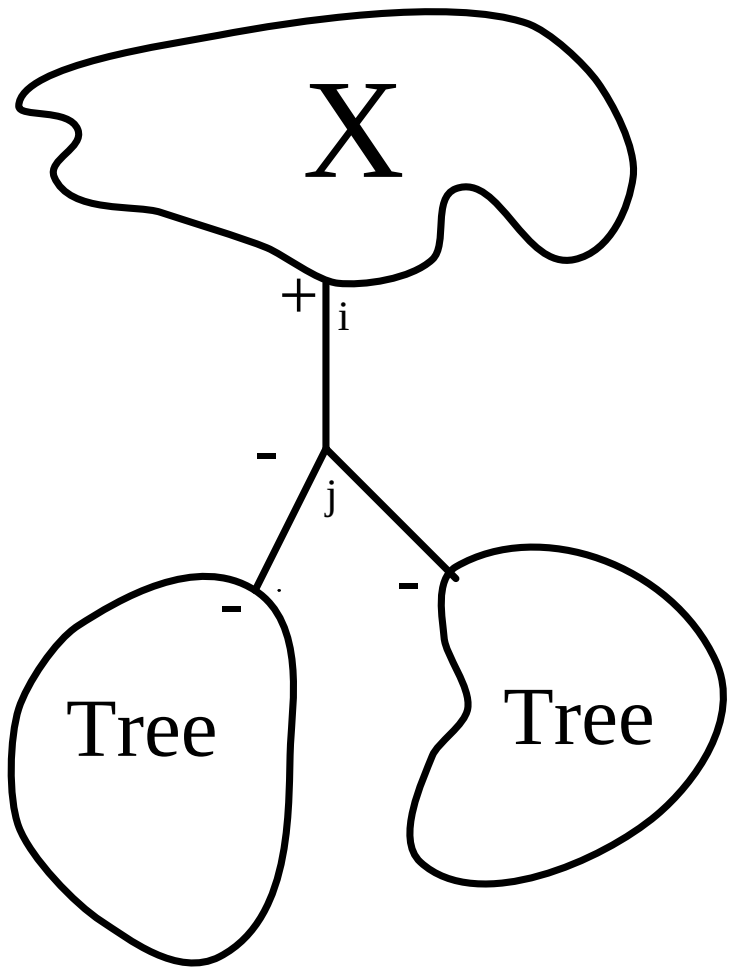

then even after insuring $s_i = +1$, $s_j$ will have 2 neighbours with $s = -1$ and will refuse flipping to $s_j = +1$. In contrast , CA will first freeze all the links belonging to the tree, delete the (i,j) link and only then will perform a flip of the obtained macro in one painless step.

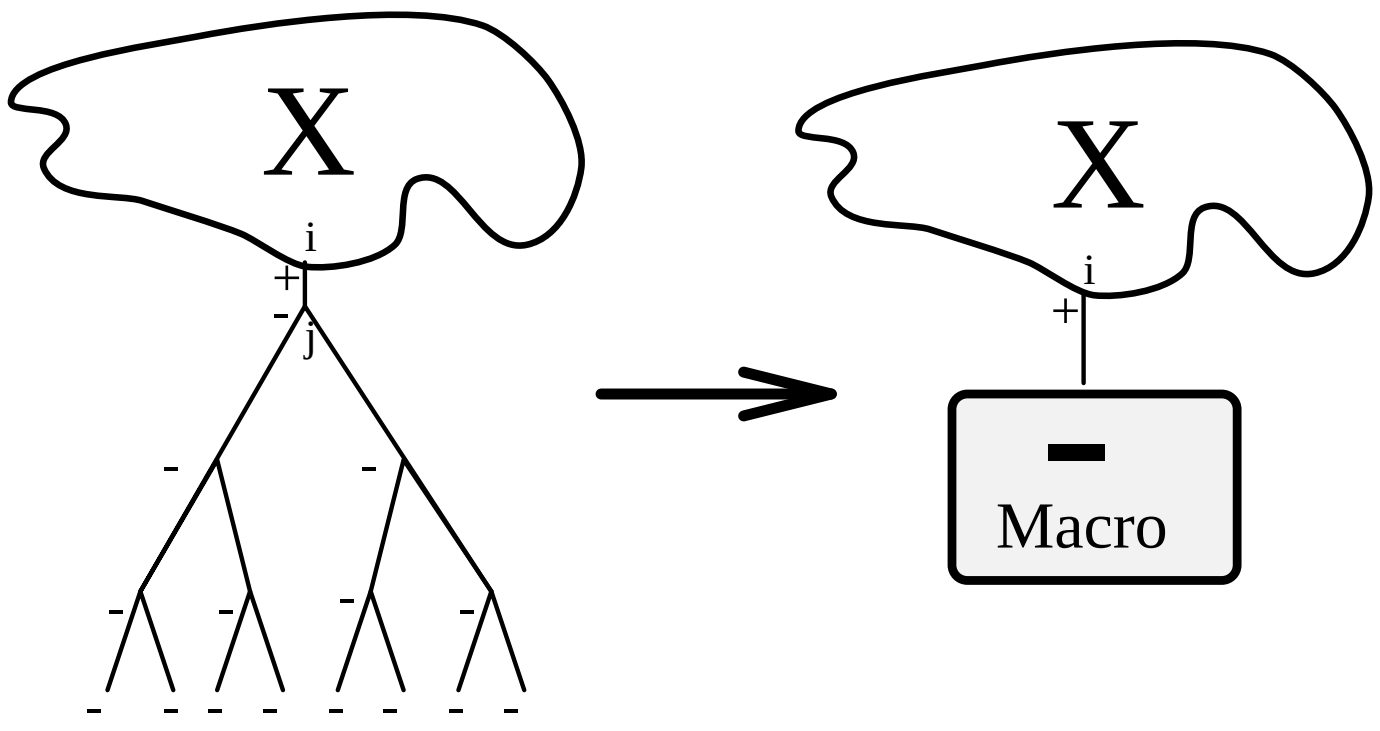

Similarly, MRA will identify each end link and trans-form its sites into a macro. Eventually, the algorithm will create a macro standing in for the entire tree. Clearly one can now update the macro of the new system in one step and then return to the explicit microscopic representation. This is similar to the DAMG procedure where the strongly connected degrees of freedom are iteratively connected into macros whose interactions are such as to represent correctly the energetics of the initial system.

### C. linear chains in gaussian distributions

Consider a configuration of the type:

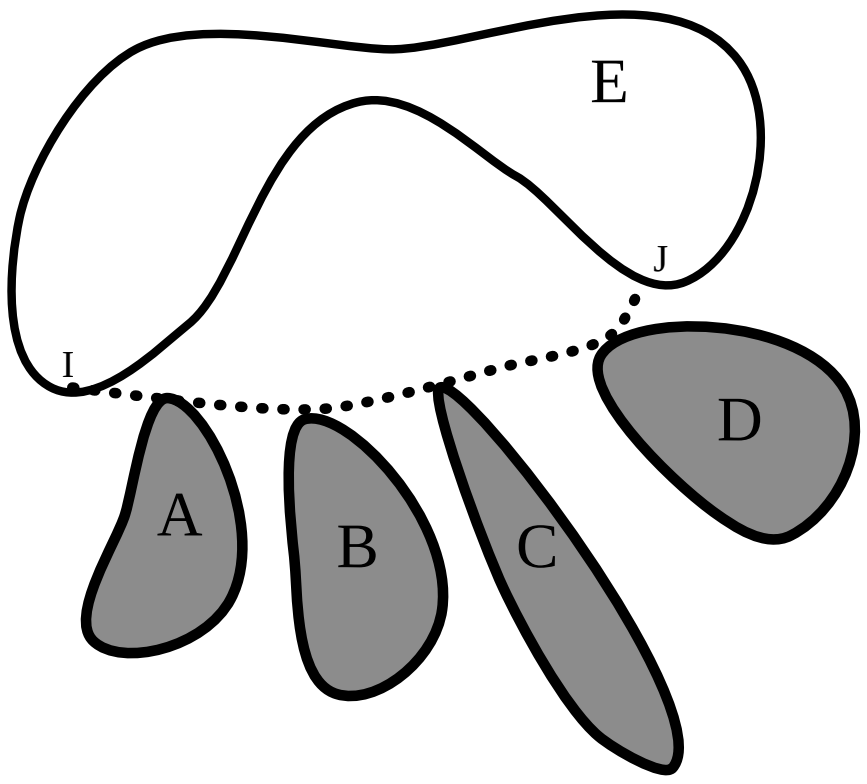

The subsets A, B , C D, (”drops”) can be considered as macros and minimized separately. Assume the block E minimization is preferring strongly a particular position for $i$ and $j$. Take it for definitiveness $s_i = +$ and $s_j = +$ and assume that the product of the links joining i to j is -. The minimal energy configuration is therefore frustrated. In order to find the minimum, one has to reach the configuration with only one unsatisfied link in the entire chain and to make such that the frustration resides on the link with the lowest $|J|$ in the chain.

As we mentioned in the previous section, CA is mainly freezing the strong satisfied links flips irreversibly the strong frustrated ones and keeps trying the weak ones. This is bringing very fast the configuration to the one with the frustration on the weakest link.

MRA (similarly to AMG) is also putting together the satisfied strong links into macros. Once the entire $\bar{ij}$ line is transformed into just one macro, (with the strength of its lowest link) one can compare the price of its frustration to the price of frustrating E (and decide which of the 2 should remain unsatisfied).

### D. Strongly coupled islands

Consider a situation in which there are islands of sites related by very strong links submerged in a sea of links which are much weaker.

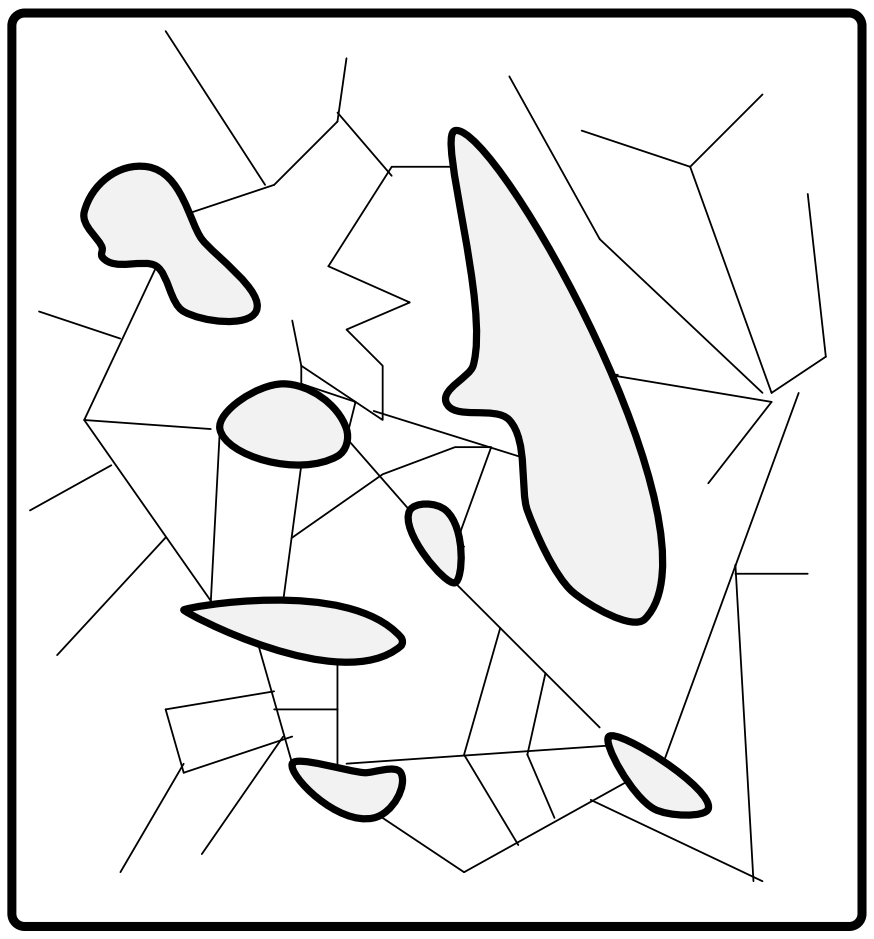

The filled zones in the figure are the islands of strong links, and we did not draw their sites explicitly [4].

Suppose the strong links are of lower density and possibly below the percolation . Suppose that the energy of the spins within the strong islands was somewhat minimized. The equilibration of the relative signs between the the various islands is very inefficient by local updating because none of the spins of an island would accept flipping without the entire island doing so. By defining each island as a macro, MRA can of course realign an entire island in one step and with quite high acceptance.

CA is efficient too in as far as it freezes the strong links and it allows the others to reach an equilibrium (especially when annealing is applied). To check this, we considered a new model (the Strong-Weak model) in which the link distribution is given by:

$$f(J) = \frac{a}{2}[\delta(J-\epsilon) + \delta(J+\epsilon)] + \frac{(1-a)}{2}[\delta(J-1) + \delta(J+1)]. \quad (7)$$

As reported in Section 6 we found indeed for the Strong-Weak model a huge gap between the satisfied energy per spin achieved by a CA and MRA on one side and by local Metropolis on the other side.

MRA is more rapid than CA but in the case in which there are more than 2 typical energies for the links, the MRA is more difficult to apply while CA is still quite efficient in implicitly identifying macros in a stochastic manner. As seen in Section 6, this is especially true when one uses an annealed schedule, in which the temperature is gradually lowered. The temperature variation enable the CA to act on clusters at various scales, corresponding to the different temperatures and to address and freeze first the stronger links into small but very strongly coupled islands. At lower energies larger, loosely coupled islands are formed and acted upon.

[4] Beware! In many aspects, this figure might be quite misleading especially in the infinite dimensional case.

## VI. COMPARING LOCAL MC TO MRA AND CA

*The Dilute SG model*

Simulations on the model defined by the equations (5) and (6) with a gaussian distribution for the links, $f(J) \propto \exp(-J^2)$, were carried out comparatively using local dynamics (Metropolis), the Cluster Algorithm (CA) [25] and the Macros Reduction Algorithm (MRA).

The simulations were carried out for various connectivity values $c$ and at temperatures below the glassy transition $T_c$. The size of the system was between 1000 and 5000 sites. The results were averaged over at least 10 different samples. A typical result is presented in Fig. 1. In Fig. 1 one sees the evolution of the energy of the system (n=5000) monitored during its computer simulation (up to 50,000 MCS per spin). One can clearly see a gap between the energies reached with the global dynamics (CA and MRA) and the local dynamics. MRA gets exactly the same energy level as the CA one but it converges much faster.

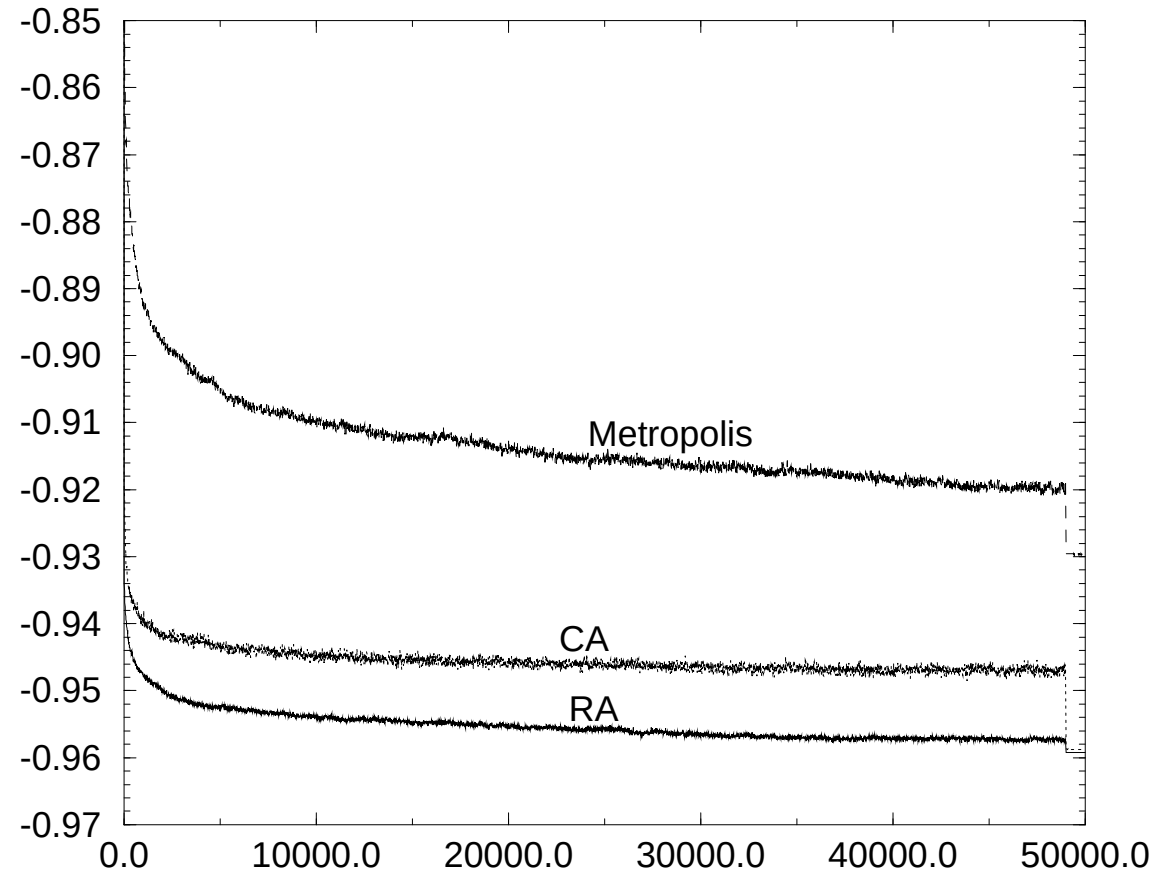

FIG. 1. $E(t)$ the energy per spin as a function of the running time for Metropolis , Cluster Algorithm (CA) and Macro Reduction Algorithm (RA) for a $c = 2, N = 5000$ lattice at $T \sim 0.3T_c$. In the last 1000 MCS we put $T = 0$.

In order to emphasize those features, we performed measurements using simulated annealing for the above dynamics. In Fig. 2 one can see similar picture of gap between the energies reached with the global dynamics (CA and MRA) and the local dynamics.

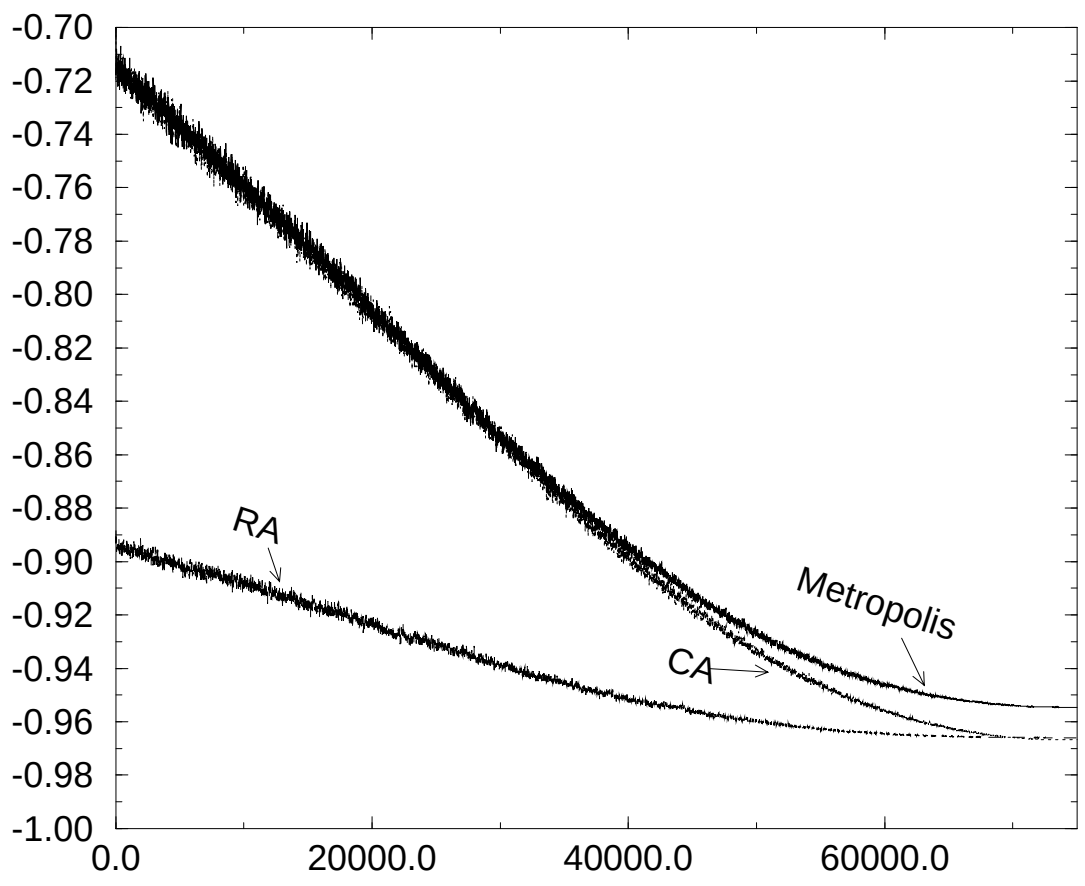

FIG. 2. Same as Fig. 1 but with an annealing schedule over the range $T \in 2.0 - 0$.

This suggests that CA and MRA reach the true vacuum while the local algorithm doesn't.

In order to check directly this issue, we look next to the GS energy as found by global and local algorithms in a model for which we could estimate it analytically.

*The Strong-Weak Model*

The simulations of the Strong-Weak model (7) were compared with the analytical results in [26]. In [26] the self-consistent description of the low temperatures of the Strong-Weak, was based on the probability distribution of the local field defined by $h_i \equiv T \tanh^{-1} < S_i >_T$. Physically, this field is the first excitation, namely, in the limit $T \to 0$, $|h_i|$ is the minimum energy cost for flipping the $i$th spin from its GS by the "best" reorganization of the system. This local field is in truth an oxymoron, since it depends on **global** properties of the cluster, the exchange field $\sum_j J_{ij} m_j$, on the other hand, is truly a **local** property depending on the local connectivity (note that $|h_i| \leq |\sum_j J_{ij} m_j|$). It was found that within the replica symmetry assumption the GS energy of the weak links is given by

$$E_W = -\frac{1}{2} c a P_0^2 \epsilon + \frac{1}{2} c(1-a)[\sum_{k=0}^{\infty}(1-4\sigma_k^2)\epsilon - \overline{h}] \quad (8)$$

where $\sigma_k = \frac{P_0}{2} + \sum_{l=1}^{k} P_l$, $\overline{h} = \sum_{h=-\infty}^{\infty} |h| P(h)$. Note that the energy of the strong links, $E_S = -\frac{1}{2}c(1-a)$, is eliminated from eq. (8).

The explicit value of $E_W$ depends on the local field, $P(h)$, which in general is difficult to calculate. Nevertheless, after some work, $P(h)$ can be determined from the equation

$$P(h) = e^{-cQ} \int_{-\infty}^{\infty} \frac{dy}{2\pi} \exp[-iyh + \frac{cQa}{2}(e^{iy\epsilon} + e^{-iy\epsilon}) + c(1-a) \sum_{l=1}^{\infty} P_l (e^{iyl\epsilon} + e^{-iyl\epsilon})] \quad (9)$$

The resulting $P(h)$ can then be compared with the usual random $J = \pm 1$ result $P_l = \exp(-cQ) I_{|l|}(cQ)$ [26] where $I_l(x)$ is the modified Bessel function. The graphs of the two $P_l$'s are presented in the insert of Fig. 3. Note that after scaling $\epsilon$ to 1, the $P_l$ for the two cases are very close. However, the values of the *exchange field* are very different:

- in the $J = \pm 1$ case the exchange field of a spin is usually not far from its local field (around the number of its neighbors), while
- in the Weak-Strong model the local field is O($\epsilon$) vs. the exchange field which is O(1).

The fact that $P_l$ is much smaller rises the hope that a global dynamics will be superior to the local one in the Strong-Weak model.

This is confirmed by the following experiments:

*Performance of Local vs. Global Algorithms in the Strong-Weak model*

In the first set of runs we choose the connectivity $c$ (the average number of neighbours) and the fraction of the strong links $(1 - a)$, such that the density of the strong links by themselves is below the percolation threshold, $c_S = c(1 - a) < 1$. It is clear that in this situation all the strong links are unfrustrated, and the frustration is located only on the weak links. Therefore only the energy of the weak links, $E_W$, is to be considered. In Fig. 3 one can see a large steady difference (35%) in the energy between the local dynamics and that of the cluster dynamics.

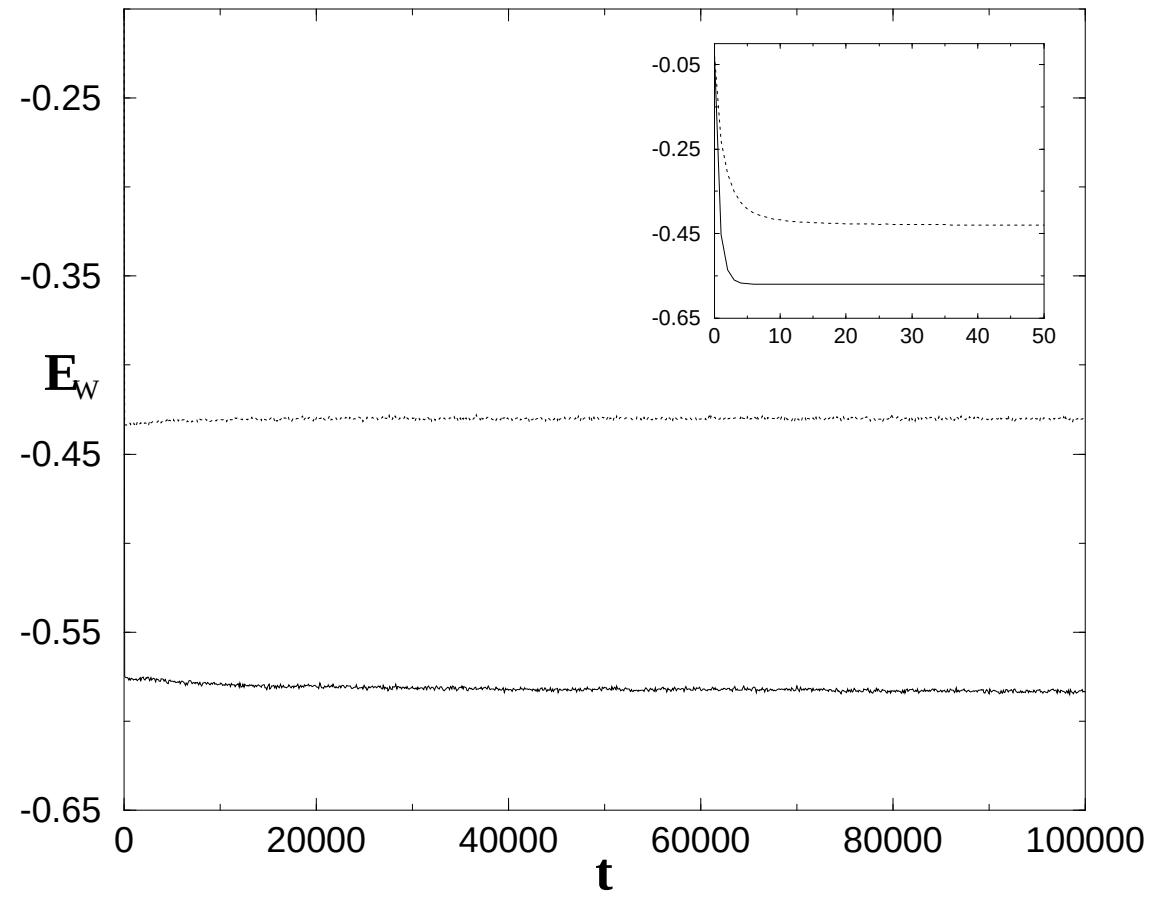

FIG. 3. Weak link energy $E_W(t)$ as a function of the running time. The model is Weak-Strong with $c = 2$, $a = 0.7$, $N = 5000$, $T = 0.5 J_W$. The strong and the weak links were scaled to 100 and 1, respectively. The solid and the dotted lines indicate Cluster and Metropolis dynamics, respectively. Inset: The first 50 steps.

*Strong-Weak with Simulated Annealing*

In the second set of runs on the Weak-Strong model, we performed measurements using simulated annealing for both global and local dynamics (FIG 4.). This was performed by cooling over a wide range of temperatures. At larger temperatures, there were the strong links which reached their minimal energy and only then, at lower temperatures, the weak links adapted to the strong links environment.

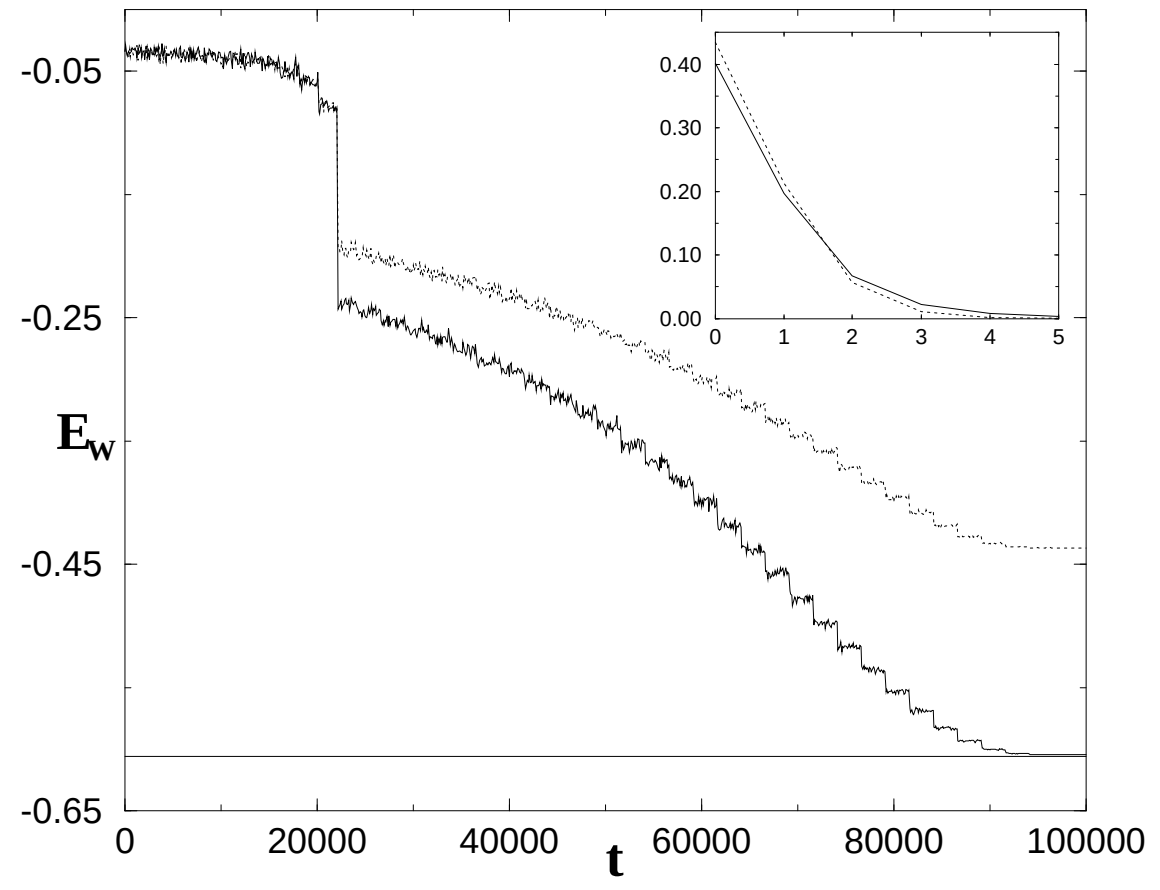

FIG. 4. $E_W(t)$ for the Weak-Strong model, $c = 2$ and $a = 0.7$. The annealing schedule range is $T \in 123 - 0$, with $\Delta T = 10$ for $T > 3$ and $\Delta T = 0.1$ for $T < 3$. The solid and the dotted lines indicate Cluster and Metropolis dynamics, respectively. The horizontal line denotes the analytical GS. Inset: $P_l$, for the WS case (solid), and $J = \pm 1$ case (dotted).

The clear energy difference between the local and cluster cases, is due to the fact that the local dynamics is totally stuck, since the probability of flipping a cluster consisting of strong links is practically zero for the local algorithm. On the other hand, the CA "knows" how to deal with the strong link structures by treating them as only "one degree of freedom", for each cluster. In other words, the CA is extremely efficient in the solving the problem: "How to arrange the weak links in the environment of the strong links".

RA gets the same results faster by defining the islands explicitly as macros and and manipulating them directly.

The type of the mean field solution for such models [26] is known to be unstable [22]. However, in Fig. 4 one can see that the analytical GS [26], is in very good agreement with the averaged GS energy obtained by the CA.

## VII. OUTLOOK

Identifying the nature and the dynamics of the macros may help understand the complex multiscale nature of a system. The techniques for identifying the relevant macros lead to a deeper understanding of the way the macroscopic description of our world arises in the study of complex systems composed of simple microscopic elements.

The macros can be multiscale reducible but in many cases there might exist complex irreducible cores. While such irreducible macros might have fortuitous characteristics, lack generality and present non-generic properties, they might be very important if the same set of cores appears recurrently in biological, neurological or cognitive systems in nature.

In such situations, rather than trying to understand the macros structure, dynamics and properties on general (multiscale, analytic) grounds as collections of their parts, one may have to recognize the unity and uniqueness of these macros and resign oneself in just making an as intimate as possible acquaintance with their features.

One may still try to treat them by the implicit elimination method [12] where the complex objects are presenting, isolating and eliminating themselves by the very fact that they are projected out by the dynamics as the slow-to-converge modes.

## ACKNOWLEDGMENTS

The research reported in this paper have been supported in part by grants from the Germany-Israel Foundation (GIF) and from the Israeli Academy of Sciences and Humanities.

This paper was completed during a visit at ICTP and SISSA. Discussions with I. Kanter, G. Mack and M. Virasoro are gratefully acknowledged.

## APPENDIX

We present in this appendix the proofs of the theorems.

### A. THEOREM I

Consider 2 subsets $A$ and $B$ of an ultrametric (UM) system $S$ which has a GS (ground state) $S^0$. Assume that the states $S^A$ and $S^B$ obtained by flipping the spins of the set $A$ (respectively $B$) in $S^0$ are GS's too.

Then

$$\#(A \bigcap B) \geq 1/2 min(\#A, \#B) \tag{10}$$

Namely, the smaller of the sets A and B share at least half of itself with the big one. Moreover, if

$$\#A > \#B$$

**then**

$$\#(A \bigcap B) = 1/2\#B. \quad (11)$$

**PROOF**:

1. $d(S^A, S^0) = \#A$, $d(S^B, S^0) = \#B$ and $d(S^B, S^A) = \#A + \#B - 2\#(A \bigcap B)$
2. On the other hand UM implies $d(S^i, S^j) \leq max[d(S^i, S^k), d(S^k, S^j)], \forall i, j, k$. Namely,
   (a) either all the distances are equal: $d(S^i, S^j) = d(S^i, S^k) = d(S^k, S^j)$ or,
   (b) two distances are equal and the third is shorter (e.g.): $d(S^i, S^j) \leq d(S^i, S^k) = d(S^k, S^j)$
3. Considering (1) and (2) for $S^0$, $S^A$ and $S^B$, one finds with some simple arithmetic that:
   (a) in the first case ($\#A \neq \#B$):
   $$\#(A \bigcap B) = 1/2 min(\#A, \#B)$$
   (b) in the second case ($\#A = \#B$):
   $$\#(A \bigcap B) \geq 1/2\#A$$

□

**Definition**
Let $A$ be a subset of the system $S$. We say $A^0$, a configuration of $A$ is **locally optimized** if $A^0$ is a minimum of the internal energy of A ignoring the energy due to the interaction with the rest of S (which we denote by $\bar{A}$).

### B. THEOREM II

**If** $S^0$ and $S^A$ differing by the sign of the spins in the region $A$ are both GS of a (not necessarily ultrametric) system S.
Denote by $A^0$ the restriction of the configuration $S^0$ to $A$.
**Then** $A^0$ is "locally optimized".

**PROOF**:

Assume that $A^0$ is **not** locally optimized therefore there exists a sub-set $A^\star \subset A$ which can be flipped to bring the system $A$ to the configurations $A^2$, which **is** a GS of $A$. Alternatively one can flip $\bar{A^\star}$ (the complement of $A^\star$ **in** $A$) to bring the system $A$ to a configurations $A^3$, which is a GS of $A$ as well.

The 2 states $A^2$ and $A^3$ differ by the flip of the entire $A$. Define the configuration $S^2$ which is $S^0$ when restricted to $\bar{A}$ and $A^2$ when restricted to $A$. Similarly define $S^3$.

The energy differences ignoring the interaction between $A$ and $\bar{A}$ are:
$e[A^0] - e[A^2] = e[A^0] - e[A^3] > O(1)$, and positive.
The overall energies are:

$$E[S^2] = e[A^2] + e[\bar{A^0}] + E[A^2 \otimes \bar{A^0}]$$

$$E[S^3] = e[A^3] + e[\bar{A^0}] + E[A^3 \otimes \bar{A^0}]$$

$$E[S^0] = e[A^0] + e[\bar{A^0}] + E[A^0 \otimes \bar{A^0}]$$

where $E[A^2 \otimes \bar{A^0}]$ is the energy of the interaction between $A^2$ and $\bar{A^0}$ etc. Using those facts together with the fact that $E[A^2 \otimes \bar{A^0}] = -E[A^3 \otimes \bar{A^0}]$, it is easy to see that:
Either

- $E[S^0] - E[S^2] > O(1)$ and positive.
  **Or**
- $E[S^0] - E[S^3] > O(1)$ and positive.

This means that at least one of the states $S^2$ or $S^3$ have lower energy than $S^0$ (by more than $O(1)$. In turn, this contradicts the initial given that $S^0$ is a GS. Therefore our assumption (that $A^0$ is not locally optimized) must be wrong, and $A^0$ **is** locally optimized.
□

### C. THEOREM III

Following the definitions of theorem II we argue:

If $A$ is locally optimized, in the GS $S^A$, and $S$ has an ultrametric structure. Then the restriction of $S^A$ to $A$ is **the unique** GS of $A$ (up to global flipping of $A$).

**PROOF**:

Following the definitions and operations at of theorem II, if the system has another GS, there exists $A^\star$ which can be flipped to bring the system $A$ to the configurations $A^2$, which is a another GS of $A$. Alternatively one can flip $\bar{A^\star}$ to bring the system $A$ to the configurations $A^3$, which is a GS of $A$ as well.
Namely:
• $e[A] - e[A^2] = e[A] - e[A^3] = O(1)$ and positive.
Using the similar reasoning as for the prove of theorem II:

- The overall energy is $E[S^A] = e[A^2] + e[\bar{A^2}] + E[A^2 \otimes \bar{A^2}]$
- $E[A^2 \otimes \bar{A^2}] = O(1)$
- This imply that $E[A^{\star 2} \otimes \bar{A^2}] = O(1)$ **and** $E[\bar{A^{\star 2}} \otimes \bar{A^2}] = O(1)$, otherwise, one of the configuration $S^2$ or $S^3$ has a lower energy (> O(1) than $S^0$).

One obtains:
• **both** $E[S^0]-E[S^2]=O(1)$ and $E[S^0]-E[S^3]=O(1)$

Therefore $A^\star$ and $\bar{A}^\star$ are two separated objects which by flipping can bring the GS $S^0$ to other 2 GS's. But this violates theorem I. Therefore the assumption that $A$ may have more than one ground state is false. □

Note that the results are independent from the definition of what is a GS. For instance in theorem II, if one would be interested in "GS-$\alpha$"-states defined by the relation $E_{S^0}-E_{S^A}\le O(N^{1/\alpha})$ (where $S^0$ is a GS). The theorem would be still valid in terms of "$\alpha$-optimization".